\documentclass[a4paper,12pt]{article}
\usepackage[lmargin=2cm,rmargin=2cm]{geometry}
\usepackage[utf8]{inputenc}
\usepackage[T1]{fontenc}
\usepackage{amssymb,amsmath,mathtools,amsthm}
\usepackage{physics}
\usepackage{listings}
\usepackage{hyperref,url,cite}
\usepackage{amsthm}
\usepackage{comment}
\usepackage{algorithm}
\usepackage[noend]{algorithmic}
\usepackage{authblk}
\usepackage{tikz}
\usetikzlibrary{quantikz,arrows.meta}
\usepackage{subcaption}
\usepackage{amsmath}
\usepackage[mode=buildnew]{standalone}
\usepackage{graphicx}

\usepackage{multirow}
\usepackage{booktabs}

\newcommand{\ii}{\textrm{i}}
\newcommand{\Hl}{\mathcal{H}}

\title{Integer Factorization through Func-QAOA}

\author[1]{Mostafa Atallah}
\author[2]{Haemanth Velmurugan}
\author[3]{Rohan Sharma}
\author[4]{Siddhant Midha}

\author[5]{Shamim Al Mamun}
\author[6,7]{Ludmila Botelho}
\author[7]{Adam Glos}
\author[7,8]{\"Ozlem Salehi\thanks{ozlemsalehi@gmail.com}}

\affil[1]{Department of Physics, Faculty of Science, Cairo University, Giza 12613, Egypt}
\affil[2]{Department of Computer Science and Engineering, NIT Tiruchirappalli, India}
\affil[3]{Department of Applied Geophysics, IIT Dhanbad, India}
\affil[4]{Department of Electrical Engineering, IIT Bombay, India}
\affil[5]{Department of Electrical and Electronics Engineering, Islamic University of Technology, Gazipur, Dhaka 1704}
\affil[6]{Joint Doctoral School, Silesian University of Technology, Akademicka 2A, Gliwice, Poland}
\affil[7]{Institute of Theoretical and Applied Informatics, Polish Academy of
	Sciences, Poland}
\affil[8]{QWorld Association, Tallinn, Estonia}
\date{}

\begin{document}

\maketitle

\begin{abstract}
Integer factorization is a significant problem, with implications for the security of widely-used cryptographic schemes. No efficient classical algorithm for polynomial-time integer factorization has been found despite extensive research. Although Peter Shor's breakthrough quantum algorithm offers a viable solution, current limitations of noisy intermediate-scale quantum (NISQ) computers hinder its practical implementation. To address this, researchers have explored alternative methods for factorization suitable for NISQ devices. One such method is the Quantum Approximate Optimization Algorithm, which treats factoring as an optimization problem defined over binary bits, resulting in various problematic aspects. In this paper, we explore the Func-QAOA approach for factorization, which premises overcoming some of the limitations of previous approaches and allows the incorporation of more advanced factorization techniques.  After reviewing the most promising quantum implementations for integer arithmetics, we present a  few illustrative examples to demonstrate the efficacy of the Func-QAOA approach and discuss methods to reduce the search space to speed up the optimization process.
\end{abstract}

\section{Introduction}
% Ozlem
Integer factorization is one of the most significant problems lying at the heart of modern cryptography. Widely used RSA cryptographic scheme \cite{rivest1978method} relies on the fact that no efficient classical algorithm is known that can be used to factor large integers in a feasible time. Despite extensive research, no polynomial-time algorithm has been discovered to solve the problem, yet it has not been definitively proven that no such algorithm exists. Classical algorithms like trial division, Fermat's factorization~\cite{lehman1974factoring}, and Pollard's rho algorithm~\cite{pollard1975monte} provide basic solutions to the problem but run in exponential time. More advanced algorithms like those employing elliptic curves~\cite{lenstra1987factoring} and general number field sieve~\cite{buhler1993factoring} offer the best complexity known so far, requiring subexponential time. 

The idea of quantum computers emerged in the 1980s, revolutionizing the field of computing by harnessing quantum phenomena to perform efficient calculations. Peter Shor made a groundbreaking discovery in 1994, putting forward a polynomial time algorithm to solve the factoring problem using quantum computers~\cite{shor1994algorithms} and recently he was awarded the Breakthrough Prize in 2022 for this achievement~\cite{prize}. The theoretical discovery was followed by experimental demonstrations of factoring number 15 using photonic and NMR qubits~\cite{vandersypen2001experimental,lu2007demonstration,politi2009shor} and improvements of the resources required for implementation~\cite{beauregard2002circuit, takahashi2006quantum, haner2016factoring}. Despite the mentioned efforts, the largest integer that one could factor using Shor's algorithm is 21~\cite{martin2012experimental} as decoherence has a destructive impact on noisy intermediate scale quantum (NISQ)~\cite{preskill2018quantum} computers which are small in size. The current estimates on the number of required physical qubits to break RSA-2048 range from millions to billions~\cite{mosca2018cybersecurity}, which is far from what we have today. 

This has led researchers to seek alternative methods for solving factorization problem suitable for NISQ. Factorization problem has been targeted in the framework of adiabatic quantum computing (AQC)~\cite{farhi2000quantum} in~\cite{peng2008quantum, schaller2007role, xu2012quantum, dattani2014quantum, peng2019factoring} and its gate-based counterpart quantum approximate optimization algorithm (QAOA)~\cite{farhi2014quantum} in~\cite{anschuetz2019variational, karamlou2021analyzing, yan2022factoring}. Those approaches require problem to be expressed as an optimization problem in the form of an Ising model. The natural cost function to minimize results from $f(x,y) = (m-xy)^2$, where $f(p,q)=0$ is the minimum of the function given that $m=pq$ is the integer to be factorized. Alternatively, one can express integers using their binary representation and end up with a set of equations that need to be satisfied, which was first observed in the context of classical optimization in~\cite{burges2002factoring}. Then, the problem Hamiltonian can be obtained by replacing binary variables with the corresponding spin variables, and further with Pauli $Z$ operators. 

For AQC, one needs to come up preferably with a 2-local problem Hamiltonian, as higher-order models are experimentally harder to implement. Some classical simplification rules and a penalty method are developed by~\cite{schaller2007role} to reduce the number of qubit requirements and to obtain a 2-local model. In~\cite{xu2012quantum}, some additional classical preprocessing rules are used to simplify the constraints, resulting in a reduction in the number of required qubits, and the integer 143 is factored using only four qubits. In~\cite{dattani2014quantum}, the authors realize that the same 4-qubit Hamiltonian can be used for factoring the integers 3599, 11663, and 56153. Other relevant research includes those using quantum annealing~\cite{jiang2018quantum, wang2020prime}, which is a heuristic optimization algorithm running in the framework of AQC~\cite{apolloni1989quantum,kadowaki1998quantum} and can be implemented on D-Wave quantum annealers~\cite{johnson2011quantum}. A 40-bit integer was recently factored using a hybrid solver of D-Wave~\cite{jun2023hubo}.

QAOA is used in the scope of factorization for the first time by Anschuetz et al.~\cite{anschuetz2019variational} under the name variational quantum factoring (VQF). A new set of classical preprocessing rules are presented, through which authors claim to reduce the number of qubit requirements to $\order{n_m}$ qubits where $n_m\sim\log m$. Building on the work of~\cite{anschuetz2019variational}, further preprocessing rules are given in \cite{karamlou2021analyzing} together with the claim that a 40-bit integer %1,099,551,473,989 
is factored using a superconducting processor consisting of only three qubits. In a recent work by Yan et al.~\cite{yan2022factoring}, QAOA is proposed to speedup Schnorr's algorithm which is based on classical lattice reduction, to obtain an algorithm that uses only $\order{n_m/ \log n_m}$ qubits. The authors report that they have successfully factored a 48-bit integer using a 10-qubit quantum processor. 

Along with these initial advancements in the field, the NISQ-era efforts for factorization have faced criticism from other researchers for various reasons. One line of criticism~\cite{fgrieustack, algasert} centers around the fact that most of the works use biprimes of special form~\cite{xu2012quantum,dattani2014quantum}, which can be factored using only a few qubits after classical preprocessing and thus cannot be used for general biprimes. Thus, the resulting binary models are so simple that they are easily tractable classically. Moreover, the idea of VQF and the claims made in~\cite{karamlou2021analyzing} and~\cite{yan2022factoring} encountered more significant criticisms, drawing reactions from various researchers, including Scott Aaronson~\cite{scott,scott2}. %While one criticism targets the QAOA algorithm itself, arguing the lack of proven speedup, 
The primary criticism is directed towards VQF's treatment of the factoring problem as an elementary optimization task reduced to a satisfiability problem, overlooking any underlying mathematical structure.
% and loses the context of factoring, leading to the conclusion that VQF can provide at best polynomial speed-up with respect to general SAT problems, which is still far away from what is offered by the current state-of-the-art classical factorization algorithms.

In this paper, we concretize and highlight additional concerns and problematic aspects associated with approaching factorization as a satisfiability problem and propose the Func-QAOA approach for factorization to mitigate these drawbacks. First of all, as a consequence of expressing the problem as binary multiplication of integers, additional variables are generated like the carry-bits~\cite{anschuetz2019variational}, even at the initial step of constructing a binary optimization model. This results in generation of redundant bits or even ones lacking interpretation with the original problem. The commonly used $X$-mixer in VQF further aggravates the problem, as it allows amplitude exchange between states which are conceptually unrelated in scope of the factoring problem. Finally, constructing the binary model from the simplistic approach of the product of two integers is unlikely to provide any speedup against classical algorithms.
% Finally, constructing the binary model from the simplistic approach of the product of two integers is unlikely to provide any speedup against classical algorithms, not only because it omits the underlying mathematical structure but also because of the large search space generated.
Taking into consideration the mentioned aspects, we describe a framework to utilize Func-QAOA for the factorization problem and analyze its potential to address and alleviate some of the aforementioned criticisms. Proposed in~\cite{bako2022near}, Func-QAOA offers an alternative implementation for QAOA that eliminates the need for an Ising model representation of the problem. Although we do not give a concrete example that provides practical advantage, building upon this novel approach, we demonstrate how the proposed framework can remedy the aforementioned issues through the ingenious design of initialization, ansatz and mixer circuits. In addition, we investigate how the context of factorization can be incorporated into the optimization process through the use of more advanced factorization techniques.

The paper is organized as follows. In Sec.~\ref{sec:preliminaries} we recall the factorization problem, QAOA and Func-QAOA. In Sec.~\ref{sec:results} we present a few pedagogical examples on how one can employ Func-QAOA framework for factorization. In the same section we concretize and highlight concerns and problematic aspects and show how the proposed framework mitigates some of them. Later we present some methods to reduce the search space. In Sec.~\ref{sec:conclusions} we discuss the results presented in the paper and conclude.

\section{Preliminaries} \label{sec:preliminaries}

\subsection{Factorization problem}

Factorization problem if considered in the context of cryptographic applications, naturally becomes a search problem: given a composite integer $m$, the goal is to find a factors $\bar{p}, \bar{q}$ s.t. their product results in $m$. Among the set of solutions $\{(p,q):1<p,q<n\}$ which we call the search space, we look for the correct solution $(\bar{p}, \bar{q})$. Given a particular solution $(p,q)$, using the definition of the integer factorization problem, it is enough to certify whether $m=pq$. One can consider different search spaces and certificates for the very same problem. Some of those include:
\begin{enumerate}
    \item Looking for $(a,b)$ s.t. $a^2-b^2 =m$, where the factors can be recovered as $\bar{p}=a-b$ and $\bar{q}=a+b$.
    \item Looking for the smallest even $r$ such that $a^r = 1\mod m$ and $a^{r/2} \neq -1\mod m$, where $a\leq m$ is an integer such that $gcd(a,m)=1$, where the factors can be computed by $\gcd(a^{r/2}\pm 1,m)$. 
    \item More advanced search spaces and certifications like those based on the elliptic curves.
\end{enumerate}
Note that the correct solution in general does not have to consist of the factors $\bar{p}$ and $\bar{q}$, but allows efficient construction of them.

Many of the algorithms targeting factorization problem work by taking a sample from the search space, which is then analysed to see whether it is a correct solution that allows to produce the factors. If this is the case, the algorithm ends, but otherwise, another sample is taken and the procedure is repeated. For this type of algorithms, we distinguish two critical properties. First one is the search space of the algorithm. Directly connected to this is the probability of obtaining the correct solution. For example, if we are sampling a candidate factor from $\{2,\dots,\lfloor \sqrt{m}\rfloor \}$, the probability of finding the correct solution is $\approx \frac{1}{\sqrt m}$. Another important property of the algorithm is the certifying time $t$ that is required for certifying the given input. Note that the total time for the algorithm to work will be the certifying time over the probability of finding the solution. 

Currently no classical algorithm is known to have a total time that is polynomial; however, the underlying reason might may vary for each algorithm. For example, for the aforementioned naive algorithm in which a potential factor $p$ is sampled, the certifying time is polynomial  -- it is enough to verify if $m = 0 \mod p$. The problem with this algorithm lies in sampling the correct solution as the corresponding probability is low. On the other hand, one can consider more advanced search algorithms instead of random sampling. This opens a room for quantum search techniques like Grover's search \cite{grover1996fast,bernstein2017post} that can accelerate classical algorithms and optimization techniques, which may provide the answer much faster than the classical counterparts.

%At the same time, there are algorithms where both the probability of finding a correct solution and the expected time for certifying the solution is large. As an example one can consider Pollard's rho algorithm, where the probability of obtaining the correct solution is high but certifying each solution requires $\order{\sqrt[4]{m}}$ iterations. Even more extreme example is Fermat's factorization method, in which there is no input (or equivalently we can introduce a single dummy input) yet the algorithm has high complexity in general (it is however very efficient if the difference of factors is small~\cite{}).

An interesting class of algorithms are those which are based on elliptic curves. These algorithms have comparatively large, inverse of subexponential probability of finding the correct solution while the certifying time is also subexponential. Hence, the total time is subexponential as well, resulting in elliptic curve factorization methods which are among the best classical algorithms known so far. This was used to provide a candidate for the Grover-based fault-tolerant quantum algorithm~\cite{bernstein2017post}.

\subsection{Quantum Approximate Optimization Algorithm}

Quantum Approximate Optimization Algorithm (QAOA)~\cite{farhi2014quantum} is a NISQ-era variational  algorithm for which the ansatz originates from the quantum adiabatic theorem~\cite{born1928beweis}. For the tunable parameters $\theta_i^{\rm mix},\theta_i^{\rm obj}$ for $i=1,\dots,m$, the optimized quantum state $| \theta^{\rm mix} ,\theta^{\rm obj}\rangle$ takes the form
\begin{equation}
    |\theta^{\rm mix},\theta^{\rm obj}\rangle = \prod_{i=1}^m \exp(-\ii \theta_i^{\rm mix} H_{\rm mix})\exp(-\ii \theta_i^{\rm obj} H)\ket{\psi_0},
\end{equation}
where $H$ is a problem Hamiltonian, a diagonal Hamiltonian whose ground state encodes the optimal solution, and $H_{\rm mix}$ is the so-called mixer Hamiltonian which is responsible for transferring the amplitude between the computational basis states. $\ket{\psi_0}$ is a fixed initial quantum state, usually (but not necessarily) a ground state of $H_{\rm mix}$. In the original QAOA, it was proposed to choose $\ket{\psi}$ as the uniform superposition and $H_{\rm mix}$ as the sum of 1-local Pauli-$X$ operators, i.e. $H_{\rm mix} = -\sum_i X_i$. However, in subsequent research, various mixers have been proposed together with suitable initial states, which allow starting in a quantum state with a better overlap with the low energy states~\cite{hadfield2019quantum, wang2020x, bartschi2020grover, sawaya2022encoding}. Owing to that development, the optimization may be performed in a much smaller space, possibly at a higher cost of the circuit depth or the number of gates. A particularly interesting is the Grover mixer~\cite{bartschi2020grover}, which requires a circuit that transforms $\ket{0}$ state into a $\ket{\psi_0}$ being a superposition of some set of solutions, and then reuses the same circuit to allow all-to-all amplitude transfer between the solutions present in $\ket{\psi_0}$. Note that the original mixer  $-\sum_i X_i$ can be in turn interpreted as transferring amplitude along the hypercube of the set of all solutions, as $X_i$ can be considered as a NOT operation acting on the $i$-th qubit.

Just as there is considerable freedom in choosing the mixer Hamiltonian and the initial state, there is flexibility in constructing the problem Hamiltonian. A typical choice is to start with a quadratic unconstrained binary optimization (QUBO) formulation which then can be transformed into 2-local Ising model. A plethora of QUBO formulations for various important paradigmatic problems were proposed in~\cite{lucas2014ising}. Subsequently, higher-order binary optimization (HOBO) formulations were proposed as a remedy for the possibly large number of qubits~\cite{glos2020space, tabi2020quantum}. Contrary to QUBO, HOBO is an arbitrary order pseudo-Boolean polynomial. In this case, it has to be guaranteed that the corresponding Ising model has a polynomial number of terms, to ensure its feasibility for execution on a quantum computer. Alternatively, for the Max-$K$-Cut and Knapsack problems,  QAOA ansatz constructions have been proposed without giving the explicit QUBO or HOBO formulation \cite{de2019knapsack, fuchs2021efficient}. The idea is generalized and formally introduced in \cite{bako2022near} under the name Func-QAOA.

FUNC-QAOA begins with an abstract description of the problem, eliminating the need for an Ising model representation. In many of the combinatorial optimization tasks, the problem involves minimization of an objective function subject to some constraints, which makes up the problem definition. For the objective, a classical program is designed to compute the objective value, and for each constraint, a classical program is designed that outputs a positive number if the constraint is not satisfied and 0 otherwise. These programs are then translated into quantum circuits and combined with a rotation around the $Z$ axis to achieve an equivalent effect to term-by-term implementation of the conventional problem Hamiltonian implementation of QAOA. The FUNC-QAOA framework enables a direct mapping from abstract problem descriptions to quantum circuits, greatly increasing the number of problem Hamiltonians that can be implemented efficiently. Furthermore, it is demonstrated that near-optimal circuits in terms of e.g. number of qubits and gates can be obtained for Travelling Salesman Problem and Max-$K$-Cut using FUNC-QAOA~\cite{bako2022near}. 

As an example, let us consider the graph coloring problem which is a decision problem asserting whether $K$ colors are sufficient to color the nodes of a particular graph $G=(V,E)$, so that adjacent nodes are colored with different colors. Since this is a decision problem, it is natural to define it as a set of constraint s.t. for each edge $\{u,v\} \in E$ we have $c(u) \neq c(v)$ where $c:V\to \{0,\dots, K-1\}$ is the coloring map. Since coloring maps form a natural search space for this problem, we require $|V|$ registers each with $\lceil \log K \rceil$ qubits so that $v$-th register with state $\ket{c(v)}$ encodes the color $c(v)$ (in binary encoding) of the node $v\in V$. Then a particular way of asserting that $c$ is a proper coloring is to follow the steps listed below for each edge $\{u,v\} \in E$:
\begin{enumerate}
    \item On register $\ket{c(v)}$ store the XOR (bit-wise addition modulo 2) of $\ket{c(v)}$ and $\ket{c(u)}$ which can be done with $\lceil \log K \rceil$ CNOTs.
    \item Apply $X$ gate (NOT) on all the qubits of $\ket{c(v)}$; if the edge connects nodes with the same color, register $\ket{c(v)}$ will be just $\ket{1\dots1}$.
    \item Implement multi-controlled NOT operation with control on $\ket{c(v)}$ and target on additional auxilliary qubit $\ket{\rm flag}$ which is initially set to $\ket{0}$.
    \item Apply $Z$-rotation on the $\ket{\rm flag}$ qubit with an magnitude proportional to the trained parameter $\theta^{\rm obj}$, which introduces the phase for the input $\ket{1}$.
    \item uncompute steps 1.-3.
\end{enumerate}
The problem Hamiltonian corresponding to the above procedure takes the form
\begin{equation}
    \sum_{\{u,v\}\in E}[c(u) = c(v)],
\end{equation}
where $[\varphi]$ is the Iverson notation, which can be used to estimate the energy of the quantum state. The function of the form above that allows to compute the energy of the measured solution is called a constraint function. Note that for the particular map $c$, the constraint function gives value 0 if and only if all pairs of adjacent nodes are coloured differently.  

\section{FUNC-QAOA for integer factorization} \label{sec:results}
In this section, we show how the FUNC-QAOA framework can lead to an interesting candidate for integer factorization. In  Sec.~\ref{sec:results-demonstration}, we demonstrate the idea through pedagogical examples. Then in Sec.~\ref{sec:results-benefits}, we analyze the problematic aspects of solving integer factorization with QAOA and discuss potential benefits achievable by using the proposed framework. Finally, in Sec.~\ref{sec:search-space-reduction}, we include technical remarks on reducing the search space.

\subsection{Demonstration of Func-QAOA through pedagogical examples} \label{sec:results-demonstration}

Designing a FUNC-QAOA circuit involves 4 main steps: Providing a problem description involving constraints and the objective function (if exists), writing classical programs for the constraints and the objective function, implementing the corresponding circuits and choosing appropriate initial state and mixer implementation. In this subsection we show examples for all the steps based on the multiplication, modulo, and order-finding certificates.

\begin{figure}[t]
    \centering
    \begin{quantikz}[column sep=0.25cm, row sep=0.7cm,]
\lstick{$\ket{\psi_p}$} & \gate[3,disable auto height]{\begin{array}{c} \text{M} \\ \text{U} \\ \text{L} \end{array}} & \qw & \qw & \qw & \gate[3,disable auto height]{\left(\begin{array}{c} \text{M} \\\text{U} \\\text{L} \end{array}\right)^{\dagger}} & \gate[3,disable auto height]{\begin{array}{c} \text{M} \\ \text{I} \\ \text{X} \\ \text{E} \\ \text{R} \end{array}}& \qw  \\
\lstick{$\ket{\psi_q}$} & & \qw & \qw & \qw & & & \qw\\
\lstick{$\ket{0}$} & & \gate{\oplus m} & \gate{ Z\textrm{-rotation}} & \gate{\oplus m} & & & \qw
\end{quantikz}
    \caption{Single layer of the Func-QAOA for multiplication-based certification.}
    
    \label{figure:func-qaoa-multiplication}
\end{figure}
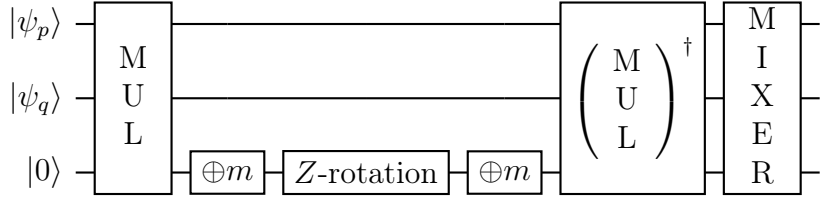

We will first consider the search space defined as all possible pairs of factors of $m$, namely $\{(p,q): 1 < p \leq \sqrt{n} \leq q \leq n/2\}$. The goal is to find the pair $(\bar p,\bar q)$ such that $m=\bar p \bar q$. For this particular problem definition, we have a single constraint $m=pq$. In the given search space, there is only one pair of integers which satisfy the constraint. The Func-QAOA algorithm works as follows. The two registers $\Hl_p$ and $\Hl_q$ are initialized to a particular initial state (this aspect will be considered later in this section). Then the constraint is implemented by taking the following steps:
\begin{enumerate}
    \item Using an arbitrary implementation of the multiplication (possibly with some auxiliary qubits), store the product of $p$ and $q$ from $\Hl_p$ and $\Hl_q$ (in computational basis) in an additional subsystem  $\Hl_{\rm product}$ which is initially set to $\ket{0\dots0}$.
    \item Modify the state of $\Hl_{\rm product}$ by taking bit-wise XOR with $m$.
    \item Apply rotation in $Z$ basis on the $\Hl_{\rm product}$ proportional to the trained parameter.
    \item Uncompute steps 1.-2. 
\end{enumerate}
The corresponding quantum circuit can be found in Fig.~\ref{figure:func-qaoa-multiplication}. The number of qubits in $\Hl_{\rm product}$ should be sufficiently large so that any product of $p$ and $q$ that can be produced can be stored.  Note that the state of $\Hl_{\rm product}$ after step 2. is equal to $\ket{0\dots 0}$ iff the computational basis states of $\Hl_p$ and $\Hl_q$ are factors of $m$.

The framework above has a few degrees of freedom. First of all, there is a plethora of multiplication algorithms one can choose. Those differ in particular in the number of auxiliary qubits, depth, number of gates, and other aspects that might be useful when considering Func-QAOA in NISQ-era computing, see Table.~\ref{tab:multiplication} for reference. The choice of the particular algorithm is irrelevant to the Func-QAOA itself and can be done based on the properties of the quantum hardware at hand. However, we can see that depending on the quantum hardware at hand, one can choose implementations optimizing the number of auxiliary qubits like the ones presented in \cite{rines2018high,gidney2019asymptotically,article}, or number of gates \cite{gidney2019asymptotically,dutta2018quantum}. Especially for the number of gates, we can observe a provable decrease in the number of CNOTs: it becomes strictly smaller than $\order{n^2}$, which is typically observed for dense QUBO formulations.

Another degree of freedom is the choice on how the phase to wrong solutions is added. For the above, a natural choice is to apply qubit-wise rotations, which would result in the phase proportional to the number of bits at which $pq$ and $m$ differ. For this Hamiltonian, the ground state matches the correct solution $(\bar p,\bar q)$. Alternatively, one can perform further operations to compute $(m-pq)^2$ or $|m-pq|$ and apply the phase proportional to that value. However in here because the numbers are exponentially large, an exponentially precise rotation would need to be used. Furthermore, this may require exponentially many shots from the quantum hardware to bring sufficient estimation of the energy. 

\newcommand{\taborder}[1]{#1}
\newcommand\Tstrut{\rule{0pt}{2.6ex}}         % = `top' strut

\begin{table}[t]\small 
\centering    
\begin{tabular}{p{4.3cm}lllp{6.3cm}}
\toprule
 & Qubits & Gates & Depth & Additional information \\
\hline \midrule
Rines et al. \cite{rines2018high} &  $\taborder{\log n}$  & $^\ast\taborder{n^2}$ & $\taborder{n}$ & Discusses Montgomery and Barrett reduction techniques; QFT-based. \\[.8ex]%\cline{2-5}
& $\taborder{n}$ & $\taborder{n^2}$ & $\taborder{n\log n}$ & Uses prefix carry lookahead adders. \\[.8ex]%\cline{2-5}
 & $\taborder{n}$ & $\taborder{n^2}$ & $\taborder{n^2}$ & Uses ripple adders \\
\midrule
Gidney \cite{gidney2019asymptotically} & $\taborder{n}$  & $\taborder{n^{\log 3}}$ & $\taborder{n^{\log 3 - 1} \log^2 n}$
 & Uses divide and conquer approach; has high overheads and is more efficient than schoolbook multiplication at around 10000 bits. \\
\midrule
Dutta et al. \cite{dutta2018quantum} & $\taborder{n^{1.404}}$ & $\taborder{n^{\log_6 16}}$ & $\taborder{n^{1.143}}$ & Toom-2.5 algorithm; performs better than naive multiplication beyond 300 bits. \\
\midrule
Larasati et al. \cite{2021}& $\taborder{n^{1.353}}$ & $\taborder{n^2}$ & $\taborder{n^{1.112}}$ & Toom-3 algorithm. \\
\midrule
Babu et al. \cite{2017QuIP...16...30B} & $\taborder{n^2}$ & $\taborder{n^2}$ & $\taborder{n}$ & Optimal tree-based multiplication technique. \\
\midrule
Álvarez-Sánchez et al. \cite{article} & $\taborder{n}$ & --- & $\taborder{\log^2 n}$ & Quantum version of Booth’s algorithm. \\
\bottomrule
\end{tabular}
    \caption{Review of quantum implementations for integer multiplication. Only the leading terms, without the constants are presented. `$\ast$' indicates that exponential-precision gates are required, which is likely not practical for NISQ-era quantum computing. Note that we refer to number of auxiliary qubits by qubits.}
    \label{tab:multiplication}
\end{table}

Finally, there is a freedom in the choice of the initial state and the corresponding mixer. We believe that particularly interesting choices rely on the Grover mixer~\cite{bartschi2020grover}. Grover mixer uses the initial state preparation circuit, to allow all-to-all transition only between the solutions that has nonzero amplitude in the initial state. Since the problem Hamiltonian (whether implemented conventionally by trotterization, or computed as in Func-QAOA) modifies only the relative phase for each solution, Grover mixer guarantees to preserve the space of the solutions that have nonzero amplitude in the initial state. As for the initial state, one can consider starting in the uniform superposition of the search space; however, creating such state might be costly. Instead, one can just start by applying Hadamard gates on all the qubits in $\Hl_p$ and $\Hl_q$, which would results in the search space $\{(p,q): 0 \leq p < 2^{\lceil \log \sqrt m\rceil}, 0 \leq q < 2^{\lceil \log (m/2) \rceil  } \}$. This will at most quadruple the size of the search space but allow much more efficient implementation of the initial state and mixer.

Note that with Func-QAOA, we were able to naturally reduce the search space from $\order{\log ^2m}$ to 
$\order{\log m}$, by avoiding the introduction of carry bit variables as done in~\cite{anschuetz2019variational}.
However, with Func-QAOA, one can further reduce the search space by exploring the alternative search space choices. Note that if one obtains the smaller prime factor $\bar p$, then obtaining the second one $\bar q = m/\bar p$ is straightforward. This observation suggests that approaching the factorization problem by searching for pairs of products is redundant, and a modulo operation could be used instead, i.e. a certification of the form $m\ {\rm MOD}\ p\equiv  0$. The search space size is defined as the logarithm of the number of computational basis vectors spanning the quantum state during the optimization. Note that if we seek for the smaller factor, the search space size is reduced from $\sim \log_2(\sqrt m) + \log_2(m/2) $ to $\sim \log(\sqrt m) = \frac{1}{2}\log m$, resulting in $\approx 66\%$ improvement in the search space size. On the other hand, this may require a more advanced circuit for implementing modulo operation, which might be more costly in the number of quantum resources. The cost of such operation is presented in Table~\ref{tab:modulo}.  While we were able to significantly reduce the search space size, the number of gates required for modulo operation brings the number of CNOTs required back to $\order{n^2}$, which is the worst-case scenario in complexity for $n$-bit QUBO formulation.

\begin{table} \small
\centering
\begin{tabular}{p{3.5cm}lllp{6.5cm}}
\toprule
& Qubits & Gates & Depth & Additional information \\
\hline \midrule
\c{S}ahin \cite{doi:10.1142/S0219749920500355}
 & $3$ & $^\ast\taborder{n^2+m^2}$ & $\taborder{n^2}$ & QFT-based approach for signed integers with $n$-bit dividend and $m$-bit divisor. \\ 
\midrule
Thapliyal et al. \cite{8691552} & $0$ & $\taborder{n^2}$ & $\taborder{n^2}$ & A general framework for division using adder and subtractor blocks. \\
\midrule
Jamal et al. \cite{6572511} & $\taborder{n^2}$ & $\taborder{n^2}$ & $\taborder{n^2}$ & Reversible divider hardware implemented on conventional and high-speed division arrays. \\
\bottomrule
\end{tabular}
    \caption{Review of quantum implementations for modulo operation. Only the leading terms, without the constants are presented. `$\ast$' indicates that exponential-precision gates are required, which is likely not practical for NISQ-era quantum computing. Note that we refer to number of auxiliary qubits by qubits.}
    \label{tab:modulo}
\end{table}

Since FUNC-QAOA relies on universality of the gate-based model, one can design more complicated search spaces than the ones considered above. For instance, in principle, FUNC-QAOA can be used for order finding, which is the main ingredient of Shor's algorithm. We will demonstrate this only for pedagogical purposes, to further clarify how FUNC-QAOA works. Let $x<m$ be a number that is coprime with $m$. The smallest integer $r$ satisfying the relation $x^r = 1 \mod m$ is called the order. Knowing the order, one can find out the factors through simple calculations, provided that the order is even and that it satisfies the relationship $x^{r/2} \neq -1 \mod m$. Note that this procedure is probabilistic as such $r$ may not exist depending on the choice of $x$ and in such a case, the procedure should be repeated with a different $x$. Since probability of choosing a valid $x$ is constant it can be done classically and we push the  difficulty of finding the correct solution $\bar r$ to the quantum part of the computation.

\begin{table}[t!] \small
    \centering
    \begin{tabular}{p{3.5cm}lllp{6.5cm}}
    \toprule
     & Qubits & Gates & Depth & Additional information \\
    \hline \midrule
    Draper \cite{draper2000addition} & $0$ & $^\ast\taborder{n^2}$ & $\taborder{n}$ & Performs modular addition; QFT-based. \\[.8ex]  %\cline{3-6}
    & $0$ & $\taborder{n \log n}$ & $\taborder{\log n}$ & Approximate QFT-based; \cite{Ruiz_Perez_2017} extends this to full (non-modular) addition.  \\  
    \midrule
    Cuccaro et al. \cite{cuccaro2004new} & $1$ & $\taborder{n}$ & $\taborder{n}$ & Ripple adder; \cite{Wang2016ImprovedQR} reduces the number of toffoli gates used with more ancilla qubits. \\ 
    \midrule
    Thapliyal et al. \cite{Thapliyal_2013} & $0$ & $\taborder{n}$ & $\taborder{n}$ & Reversible ripple carry adder. \\ 
    \midrule
    Draper et al. \cite{10.5555/2012086.2012090} & $\taborder{n}$ & $\taborder{n}$ & $\taborder{\log n}$ & Carry lookahead adder. \\ 
    \midrule
    Thapliyal et al. \cite{inbook} & $\taborder{n}$ & $\taborder{n}$ & $\taborder{\log n}$ & Reversible carry lookahead adder; claims slight improvement over \cite{10.5555/2012086.2012090} in terms of number of gates and depth. \\ 
    \midrule
    Takahashi et al. \cite{10.5555/2011464.2011476} & $\taborder{n/d(n)}$ & $\taborder{n}$ & $\taborder{d(n)}$ & $d(n) = \Omega (\log n)$; Combination of ripple and carry lookahead adder approaches \cite{Takahashi2008AFQ}; the circuit can be transformed for linear nearest neighbor architecture without increasing the size or depth asymptotically;  also discusses improvements when unbounded fanout gates are available. \\
    \midrule
    Gidney \cite{gidney2020quantum} & $\taborder{n/b}$ & $\taborder{n+n/b-b}$ & $\taborder{b + \log n/b}$ & Block lookahead adders; parallelizes across blocks of size $b$, instead of all bits. \\ 
    \midrule
    Choi et al. \cite{Choi_2012} & $2n - \sqrt{n}$ & --- & $\taborder{\sqrt{n}}$ & Adder for 2D NTC architecture. \\
\bottomrule

\end{tabular}
    \caption{Review of quantum implementations for integer addition. Only the leading terms, without the constants are presented. `$\ast$' indicates that exponential-precision gates are required, which is likely not practical for NISQ-era quantum computing. Note that we refer to number of auxiliary qubits by qubits.}
    \label{tab:adder}
\end{table}

\begin{table}[] \small
    \centering
    \begin{tabular}{p{3.5cm}lllp{6.5cm}}
    \toprule [1pt]
     & Qubits & Gates & Depth & Additional information \\
    \hline \midrule
     Şahin \cite{doi:10.1142/S0219749920500355}& $0$ & $^\ast\taborder{n^2 + nm - m^2}$ & $\taborder{n^2}$ & QFT-based approach for subtracting $m$-bit signed integer from an $n$-bit signed integer.\\ 
     \midrule
     Thapliyal \cite{Thapliyal2016} & $n-1$ & $^\ast\taborder{n}$ & $\taborder{n}$ & Extension of half and full subtractor in \cite{thapliyal2011new}; reversible subtractor. \\[.8ex]
     & $0$ & $\taborder{n}$ & $\taborder{n}$ & Subtractor based on adder; also discusses the design of adder-subtractors. \\
\bottomrule
\end{tabular}
    \caption{Review of quantum implementations for integer subtraction. Only the leading terms, without the constants are presented. `$\ast$' indicates that exponential-precision gates are required, which is likely not practical for NISQ-era quantum computing. Note that we refer to number of auxiliary qubits by qubits.}
    \label{tab:subtractor}
\end{table}

In this case, the search space consists of integers $\{r: 2\leq r < m, r\textrm{ even}\}$ and we implement the constraint $x^r = 1 \mod m$. Note that we have an optimization problem, as we aim to find the smallest $r$. To design a FUNC-QAOA, we initialize register $\Hl_r$ to store $\ket{r}$ in equal superposition of all basis states and we choose a random $x$. The last qubit should be fixed to $\ket{0}$ as we are interested in even $r$ only. We reserve another register $\Hl_o$ to store $\ket{x^r \mod m}$ where each qubit is initialized to $\ket{0}$. Next, we apply modular exponentiation circuit to store $\ket{x^r \mod m}$. Note that various circuit constructions appear in the literature for performing modular exponentiation \cite{takahashi2006quantum, beauregard2002circuit, haner2016factoring} with trade-off between the number of qubits and the number of gates. We check if the result is equal to 1 and set an additional $\ket{flag}$ qubit to $\ket{1}$ if this is not the case. We apply rotation on $\ket{flag}$ proportional to the trained parameter. In addition, we apply rotation on $\ket{r}$ proportional to the magnitude of $r$, as we would like to minimize $r$. Finally, all the steps are uncomputed. Unfortunately, $r=0$ is a trivial answer but additional energy can be incorporated for this specific case.

Note that the above proposals are agnostic to the actual implementation of the arithmetic operations. This is because Func-QAOA is indifferent to how the operations are executed, yet it would be desirable to have a noise-robust implementations to the greatest extent possible. Therefore, the specific implementation can be chosen based on the properties of the quantum hardware at hand, taking into account aspects like limited 
connectivity, number of qubits or quantum volume. To complement this subsection, we provide a review of existing implementations of various arithmetic operations, including addition as presented in Table~\ref{tab:adder}, subtraction as presented Table~\ref{tab:subtractor}, in multiplication as presented in Table~\ref{tab:multiplication}, and modulo as presented in Table~\ref{tab:modulo}.

% General remarks and concerns on QAOA and Func-QAOA
% Remarks and concerns on QAOA and Func-QAOA
% Remarks and concerns on QAOA
% Concerns on QAOA and their mitigation
% Concerns on factoring through QAOA and their mitigation
% General remarks on applicability of Func-QAOA
% Addressing the concerns on factorization using QAOA
\subsection{Addressing the concerns on QAOA for factorization} \label{sec:results-benefits}

One of the main concerns regarding QAOA when used for solving the factorization problem is that the optimization algorithms are not suitable for the search or decision problems. Indeed, such algorithms need to find the \textit{global optimum} of the objective function to provide the correct answer, which is not always guaranteed by heuristic algorithms. The QAOA ansatz is derived from the trotterization of the adiabatic evolution, allowing to reach global optimum with a sufficiently large numbers of layers, at least with parameters chosen to be sufficiently small numbers, say $\frac{1}{l}$ for $l$ layers. Unfortunately, expecting such a convergence with a small number (thus NISQ-friendly) of layers is not obvious. Nevertheless, it is shown that QAOA outperforms the best classical solver for a particular decision problem in a recent work~\cite{boulebnane2022solving}, providing a positive argument for its application in certain scenarios. For completeness, we would like to point out a few studies on QAOA. In \cite{streif2019comparison}, the authors show the existence of instances for which quantum and simulated annealing have exponentially small probability to find the solution, while those instances are exactly solvable by QAOA. On the other hand, in another recent work \cite{pelofske2023quantum}, it is shown that QA outperforms QAOA for randomly generated Ising instance when executed on the currently available hardware. Taking these into account, it remains inconclusive whether QAOA is a viable candidate for the integer factorization problem. 

Another concern raised is that solving integer factorization problem with QAOA requires a necessary step of transforming the problem into a SAT instance, thereby concealing the original interpretation behind a new formulation. This becomes particularly evident when attempting to approach the higher-order binary optimization formulation (like the one presented in~\cite{anschuetz2019variational}) using quantum annealing -- in that case a quadratization procedure is necessary, which at the cost of introducing many additional variables, transforms the original pseudo-Boolean polynomial into a \textit{quadratic} pseudo-Boolean polynomial. However, this hinders exploiting the original formulation of the problem, reducing (as mentioned in~\cite{scott} in the comments section) the original problem to a SAT instance, for which the complexity is unlikely to be better than $c^{\log m}$ for any $c>1$. On the contrary, for QAOA, no-meaning variables like the bits coming from the quadratization are not necessary, as in the gate-based machines higher-order terms can be easily implemented. Finally, such bits will be definitely not present when formulating the objective function via the procedure for Func-QAOA.

Even though such bits do not exist in the model, there might be some other issues which `strengthen' the SAT-opted optimization evolution. The first example is the choice of the $X$-mixer, in which the amplitude-transfer graph represents a hyper-cube, preferring single-bit change. In order to avoid this, an all-to-all mixer should be used or nearest in modulo numbers should be swapped. The first one can be easily implemented with Grover mixer~\cite{bartschi2020grover}, while some efforts towards the second one can be found in~\cite{qiang2016efficient}, where the authors use the fact the corresponding mixer is a circulant matrix. Both mixers contribute to moving away from the interpretation of the landscape as a SAT instance.

Finally, perhaps the most serious issue with QAOA and also FUNC-QAOA is the choice of appropriate certificates. In order to demonstrate any advantage over the best classical algorithm, QAOA has to run in subexponential time. However, if the success probability is $\order{1/c^n}$ for some $c>1$, it is unlikely to expect a super-polynomial improvement. Note that in all the Func-QAOA examples provided earlier, like multiplication, modulo and order finding, the success probability was exponentially small, making them not so interesting candidates. Better candidates are likely to be found among the elliptic curve factorization methods. Note that Func-QAOA might prove to be particularly beneficial for such certificates, as it is easier to design complicated certificates using universality of the gate-based machines. However, the oracles for these algorithms require subexponentially many gates, which make them impractical for the NISQ-era. Despite our efforts, we have not been able to find any suitable certificate that would simultaneously require only a polynomial number of gates and qubits, while also having a subexponential gate count. 

\subsection{Search space reduction} \label{sec:search-space-reduction}

The previously presented advantages of Func-QAOA mitigate the fundamental issues arising from using QAOA-like algorithms for integer factorization. In this section, we explore how the search space can be reduced further by means of careful state preparation and Grover Mixer choice. Originally, for QAOA with the $X$-mixer, all the solutions were considered as the initial state is the superposition of all states. However in Quantum Alternating Operator Ansatz~\cite{hadfield2019quantum}, and later in \cite{wang2020x, bartschi2020grover}, alternative initial states are proposed, namely the superposition of one-hot vectors and the superposition of permutations. Reducing the search space size has a significant impact on the optimization quality, as it was numerically observed in~\cite{bako2022near}. In this section, as a pedagogical example, we will consider multiplication-based variant, however our results can be naturally considered for any other variant of the problem.

Let us first recall VQF~\cite{anschuetz2019variational}. In there, three different groups of bits are introduced: $\{p_i\}_i$ forming a register for the factor $p$, $\{q_j\}_j$ forming a register for the other factor $q$, and finally $\{z_k\}_k$ which are the carry-out bits coming from the grade-school multiplication of $\{p_i\}_i$ and $\{q_j\}_j$. The carry-outs bits are necessary to certify whether the product of $p$ and $q$ is equal to the biprime $m$. Since the problem is formulated as a Higher-Order Binary Optimization (HOBO) with $X$-mixer, $z_k$ bring additional degrees of freedom, while not introducing any information about the problem: for any values of $p$ and $q$, there is a unique natural assignment that can be obtained by taking the true carry-outs occurring in the multiplication. Recall that by using Func-QAOA, we were able to reduce the search space size from $\order{\log^2 m}$ into provable $\order{\log m}$ without any heuristic preprocessing of the problem. This was achieved by avoiding the carry bits, dropping the $\{q_j\}_j$ bits from the search space and optimizing only over the $\{p_i\}_i$ bits by choosing modulo certificate instead of multiplication. 

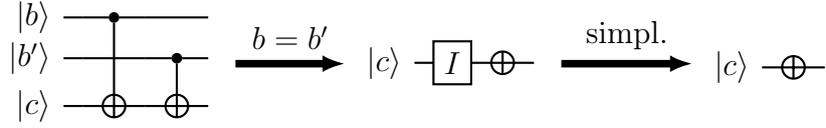
\begin{figure}
    \centering
    \begin{tikzpicture}
\node (T1) at (0,0) {
        \begin{quantikz}[column sep=0.25cm, row sep=0.4cm,]
            \lstick{$\ket{b}$} & \qw & \ctrl{1} & \qw  & \qw & \qw \\
            \lstick{$\ket{b'}$}  & \qw & \qw \vqw{1} & \qw & \ctrl{1} & \qw \\
            \lstick{$\ket{c}$} & \qw & \targ{1} & \qw & \targ{1} & \qw \\
        \end{quantikz}
};

\node (T2) at (4.5,0) {\begin{quantikz}[column sep=0.25cm, row sep=0.25cm,]
             \lstick{$\ket{c}$} & \gate{I} & \targ{} & \qw \\
         \end{quantikz}};

\node (T3) at (8.7,0){\begin{quantikz}[column sep=0.25cm, row sep=0.25cm,]
             \lstick{$\ket{c}$} &  \targ{} & \qw \\
         \end{quantikz}};;

\draw [line width=1mm, -{Latex[length=3mm]}] ($(T1.east) + (0.2,0.15)$) -- node[above] {$b = b'$} ($(T2.west) + (0,0.15)$);
\draw [line width=1mm, -{Latex[length=3mm]}] ($(T2.east) + (0.2,0.15)$) -- node[above] {simpl.} ($(T3.west) + (0.,0.15)$);

\end{tikzpicture}

%     \begin{subfigure}[t]{0.4\textwidth}
%         \centering
%         \begin{quantikz}[column sep=0.25cm, row sep=0.4cm,]
%             \lstick{$\ket{b}$} & \qw & \ctrl{1} & \qw  & \qw & \qw \\
%             \lstick{$\ket{b'}$}  & \qw & \qw \vqw{1} & \qw & \ctrl{1} & \qw \\
%             \lstick{$\ket{c}$} & \qw & \targ{1} & \qw & \targ{1} & \qw \\
%         \end{quantikz}
%         \label{fig:original_circuit}
%     \end{subfigure}
%     \begin{subfigure}[m]{0.1\textwidth}
%         \centering
%         $b=0$\\
%         {\large$\mbox{$\Longrightarrow$}$}\\
%         $b'=1$\\
%     \end{subfigure}
%     \begin{subfigure}[t]{0.4\textwidth}
%         \centering
%         \begin{quantikz}[column sep=0.25cm, row sep=0.25cm,]
%             \lstick{$\ket{c}$} & \gate{I} & \targ{} & \qw \\
%         \end{quantikz}
%         \label{fig:modified_circuit}
%     \end{subfigure}
%     \begin{subfigure}[t]{0.4\textwidth}
%         \centering
%         \begin{quantikz}[column sep=0.25cm, row sep=0.4cm,]
%             \lstick{$\ket{b}$} & \qw & \ctrl{1} & \qw  & \qw & \qw \\
%             \lstick{$\ket{b'}$}  & \qw & \qw \vqw{1} & \qw & \ctrl{1} & \qw \\
%             \lstick{$\ket{c}$} & \qw & \targ{1} & \qw & \targ{1} & \qw \\
%         \end{quantikz}
%         \label{fig:original_circuit}
%     \end{subfigure}
%     \begin{subfigure}[m]{0.1\textwidth}
%         \centering
%         $b=0$\\
%         {\large$\mbox{$\Longrightarrow$}$}\\
%         $b'=1$\\
%     \end{subfigure}
%     \begin{subfigure}[t]{0.4\textwidth}
%         \centering
%         \begin{quantikz}[column sep=0.25cm, row sep=0.25cm,]
%             \lstick{$\ket{c}$} & \gate{I} & \targ{} & \qw \\
%         \end{quantikz}
%         \label{fig:modified_circuit}
%     \end{subfigure}
    \caption{Circuit simplification by bit substitution.}
    \label{fig:bit-subs}
\end{figure}

\begin{figure}
    \centering
    \begin{tikzpicture}
\node (T1) at (0,0) {\begin{quantikz}[column sep=0.25cm, row sep=0.4cm]
                  \lstick{$\ket{b}$} & \ctrl{2}  & \qw & \ctrl{1} & \qw  \\
                  \lstick{$\ket{b'}$}   & \qw  & \ctrl{1} & \ctrl{1} & \qw  \\
                  \lstick{$\ket{c}$}  & \targ{1} & \targ{1} & \targ{1} & \qw \\
                \end{quantikz}};

\node (T2) at (4.5,0) {\begin{quantikz}[column sep=0.25cm, row sep=0.4cm,]
                    \lstick{$\ket{b'}$}  & \ctrl{1} & \ctrl{1} & \ctrl{1} & \qw \\
                    \lstick{$\ket{c}$}  & \targ{1} & \targ{1} & \targ{1} & \qw \\
                \end{quantikz}};

\node (T3) at (8.7,0) {\begin{quantikz}[column sep=0.25cm, row sep=0.4cm,]
                    \lstick{$\ket{b'}$}  & \ctrl{1} & \qw \\
                    \lstick{$\ket{c}$}  & \targ{1} & \qw \\
                \end{quantikz}};

\draw [line width=1mm, -{Latex[length=3mm]}] ($(T1.east) + (0.2,0)$) -- node[above] {$b = b'$} (T2);
\draw [line width=1mm, -{Latex[length=3mm]}] ($(T2.east) + (0.2,0)$) -- node[above] {simpl.} (T3);

\node (B1) at (-1,-2.5) {\begin{quantikz}[column sep=0.25cm, row sep=0.4cm,]
                        \lstick{$\ket{b}$} & \ctrl{2}  & \qw & \ctrl{1} & \qw  \\
                        \lstick{$\ket{b'}$}   & \qw  & \ctrl{1} & \ctrl{1} & \qw  \\
                        \lstick{$\ket{c}$}  & \targ{1} & \targ{1} & \targ{1} & \qw \\
                    \end{quantikz}};

\node (B2) at (4.5,-2.5) {\begin{quantikz}[column sep=0.25cm, row sep=0.25cm,]
            \lstick{$\ket{b'}$}   & \targ{1}  & \ctrl{1} & \targ{1}  & \ctrl{1} & \qw & \qw \\
            \lstick{$\ket{c}$}  & \qw  & \targ{1} & \qw  & \targ{1} & \gate{I} & \qw \\
        \end{quantikz}};

\node (B3) at (9.5,-2.5) {\begin{quantikz}[column sep=0.25cm, row sep=0.4cm,]
                    \lstick{$\ket{b'}$}  & \qw & \qw \\
                    \lstick{$\ket{c}$}  & \targ{1} & \qw \\
                \end{quantikz}};

\draw [line width=1mm, -{Latex[length=3mm]}] ($(B1.east) + (0.2,0)$) -- node[above] {$b = 1-b'$} (B2);
\draw [line width=1mm, -{Latex[length=3mm]}] ($(B2.east) + (0.2,0)$) -- node[above] {simpl.} (B3);

\end{tikzpicture}

    \caption{Circuit Simplification by bit redundancy.}
    \label{fig:bit-redundancy}
\end{figure}

While the above is simple and widely applicable reduction in Func-QAOA framework~\cite{bako2022near}, one can save additional resources by carefully modifying the initial state. Note that in~\cite{anschuetz2019variational}, the quadratic constraints are used to generate simplification rules, through which the original HOBO can be simplified during the classical preprocessing. For the proposed QAOA, the preprocessing was used to remove some qubits and the corresponding gates. For Func-QAOA, we use the clauses generated in VQF to simplify the initial state. We consider three scenarios:
\begin{enumerate}
    \item \emph{Bit substitution}: clauses imply $b=0$ or $b=1$ for a binary variable $b$.
    \item \emph{Bit redundancy}: clauses imply $b=b'$ or $b=1-b'$ for binary variables $b,b'$.
    \item \emph{Superposition reduction}: clauses imply binary variables $b_1,\dots,b_k$ can attain only particular values.
\end{enumerate}
For each case, we choose a unique strategy for simplifying the circuit. The simplest case is the bit substitution, where the bits provide no information from the context of the optimization and should be set to $\ket{0}$ or $\ket{1}$ for the whole process. However, this means that those qubits can be just removed from the circuit, and all the gates applied afterwards can be simplified accordingly, as presented in Fig.~\ref{fig:bit-subs}.As an example, for the multiplication, it is reasonable to assume that the least significant bits of the potential factors are 1 -- otherwise the biprime would need to be even, which is unrealistic for applications in cryptography.

For the bit redundancy, one can start with the Bell states $\frac{1}{\sqrt 2}(\ket{00} + \ket{11})$ and $\frac{1}{\sqrt 2}(\ket{01} + \ket{10})$ for constraints $b=b'$ and $b=1-b'$ respectively. This minor modification requires only a tiny increase in the number of gates, one CNOT for the state preparation and two for the Grover mixer. However, it effectively halves the size of the search space. In the case one of the bits serves solely as a control, the bit can be removed from the circuit and the gates where $b$ acts as a control can be implemented as follows. W.l.o.g., suppose we remove $b'$. If $b=b'$, then the gates controlled on $b'$ can be controlled by $b$. If $b=1-b'$, first a NOT gate should be applied on $b$, then the gate is applied with a control on $b$ and NOT gate should applied again to revert $b$ back to its original state. Through this simplification, the search space size is halved and one qubit is saved. Those examples are visualized in Fig.~\ref{fig:bit-redundancy}. 
 
While very effective, we do not expect the above simplifications to appear too frequently. On the contrary, it is much more likely that more complicated clause simplifications can be found, in which for a subset of bits only some possible assignments are valid. Here, a dedicated method is to create a superposition of only those computational basis states that satisfy the clause. If the number of qubits over which the clause is defined is sufficiently small, say at most $c\log(n)$, one can use the brute-force method to determine all bit assignments satisfying the clause and then prepare a superposition using a general state preparation method~\cite{shende2005synthesis, mottonen2005transformation} using $\order{n^c}$ gates. 

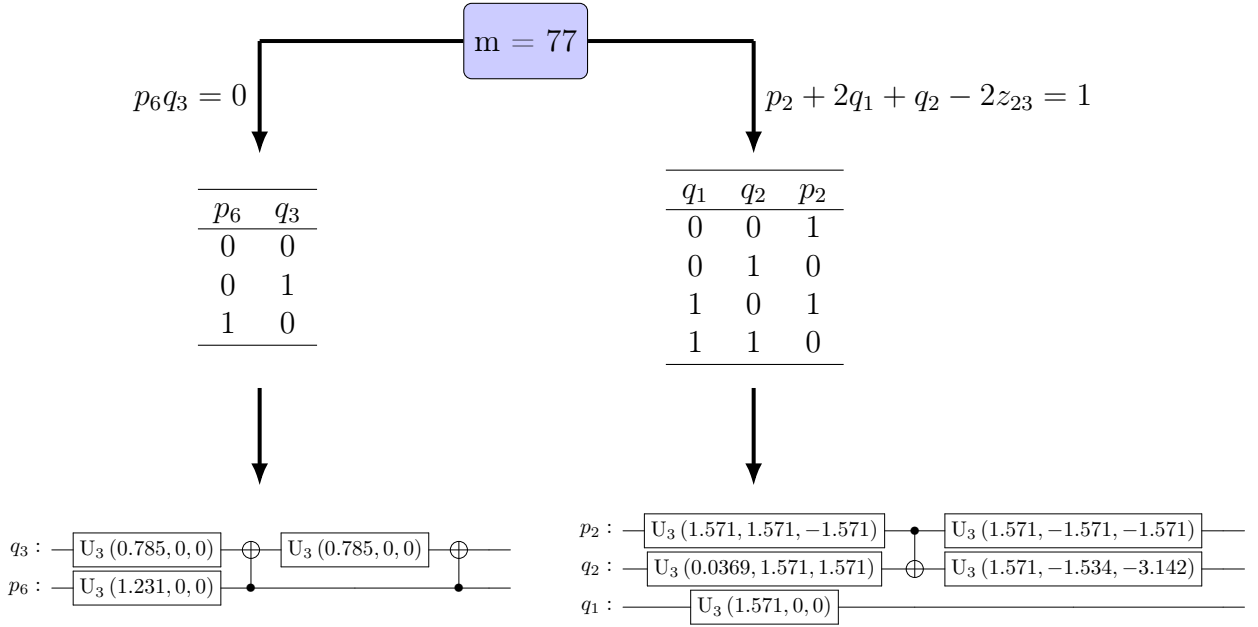
\begin{figure}
    \begin{tikzpicture}
  % Define box style
  \tikzstyle{mybox} = [draw, rectangle, minimum size=1 cm, align=center, fill=blue!20, rounded corners=3pt];
  
  % Draw the box with the number inside
  \node[mybox] (box) at (0, 0) {m = 77};
  
  % Draw arrows going to the right and curving downward (without arrow tips)
  \draw[-, line width=1.5pt] (box.east) -- (3, 0) coordinate (right);
  
  % Draw arrows going to the left and curving downward (without arrow tips)
  \draw[-, line width=1.5pt] (box.west) -- (-3.5, 0) coordinate (left);
  
  % Draw vertical arrows from the curved points (right arrow)
  \draw[-{Latex[length=3mm]}, line width=1.5pt] (3, 0) -- (3, -1.5) node[midway, right] {$p_2 + 2q_1 + q_2 - 2z_{23} =1$};
  \draw (1.7, -3) node[right] {%
    \begin{tabular}{ccc}
    \hline
      $q_1$ & $q_2$ & $p_2$ \\
      \hline
      0     & 0     & 1 \\
      0     & 1     & 0 \\
      1     & 0     & 1 \\
      1     & 1     & 0 \\
      \hline
    \end{tabular}
  };
  
  % Draw vertical arrows from the curved points (left arrow)
  \draw[-{Latex[length=3mm]}, line width=1.5pt] (-3.5, 0) -- (-3.5, -1.5) node[midway, left] {$p_6 q_3=0$};
  \draw (-2.6, -3) node[left] {%
    \begin{tabular}{cc}
    \hline
      $p_6$ & $q_3$ \\
      \hline
      0     & 0 \\
      0     & 1 \\
      1     & 0 \\
      \hline
    \end{tabular}
  };

    \draw[-{Latex[length=3mm]}, line width=1.5pt] (3, -4.6) -- (3, -5.9) node[midway, right]{};
    \draw[-{Latex[length=3mm]}, line width=1.5pt] (-3.5, -4.6) -- (-3.5, -5.9) node[midway, left]{};

    \draw (0, -7) node[right] {%
        \includegraphics[scale=0.8]{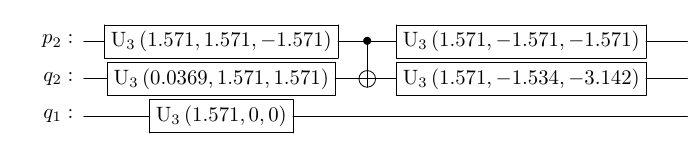}
  };

  \draw (-0., -7) node[left] {%
    \includegraphics[scale=0.8]{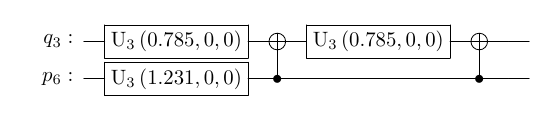}
  };

\end{tikzpicture}
    \caption{The search space reduction for $m=77$ by superposition reduction.}
    \label{fig:space-reduction-scheme}
\end{figure}

Based on the VQF formulation, we show how the search space can be reduced for $m=77$, which is demonstrated in Fig.~\ref{fig:space-reduction-scheme}. First we identify two clauses with non-overlapping variables $p_i$ and $q_j$; $p_6q_3=0$ and $p_2+2q_1+q_2-2z_{23}=1$. Then we identify all the valid assignments for these bits. Note that for the second clause, we look for an assignment of $p_2,q_1,q_2$ so that there \textit{exists} a value for $z_{23}$ and the clause is satisfied. The valid assignments are presented in Fig.~\ref{fig:space-reduction-scheme}. Then, quantum circuits are generated which will produce a uniform superposition of valid assignments for the corresponding qubits, for instance $\frac{1}{\sqrt 3}(\ket{00}+\ket{10}+\ket{01})$ for the variables $p_6,q_3$. The chosen clauses allow reducing the number of possible solutions from 4 to 3 and from 8 to 4, respectively. Consequently, the first simplification removes 25\% of the search space, while the second one removes 50\% of the search space.

Note that removing even a minimal number of bits may have a significant effect on the reduction of the effective space size. To see this, let us assume an optimistic scenario with $n$ qubits and $n/c$ clauses, each defined on a fixed, $c$ independent bits. Suppose $b<2^c$ bits are removed from each clause. Then the new search space size is
\begin{equation}
    \log_2 \prod_{i=1}^{n/c} (2^c - b) = n\left (1+\frac{1}{c}\log_2\left (1-\frac{b}{2^c}\right)\right ),
\end{equation}
whereas the original search space size is $n$. We can see that for any fixed $b$ and $c$, we are reducing the effective space size by a multiplicative constant. As an example, if we take $b=\frac{1}{2} 2^c$, we have the effective space size $n(1-\frac{1}{c})$. Note that being able to find $n/c$ clauses on non-overlapping variables might in general be a challenging task, e.g. in VQF formulation most of the clauses act on most of the binary variables instead of a tiny subset of them.

\begin{figure}
    \centering
    \begin{subfigure}{0.45\textwidth}
        \includegraphics[width=\linewidth]{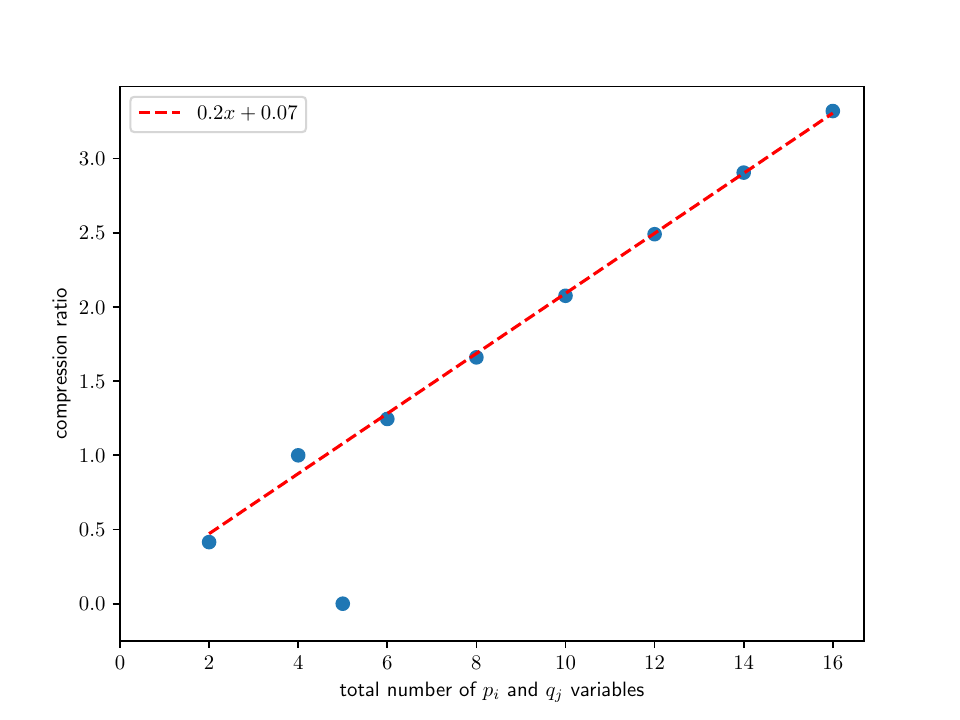}
        \caption{$m = 1687927$}
        \label{subfig:image1}
    \end{subfigure}
    \hfill
    \begin{subfigure}{0.45\textwidth}
        \includegraphics[width=\linewidth]{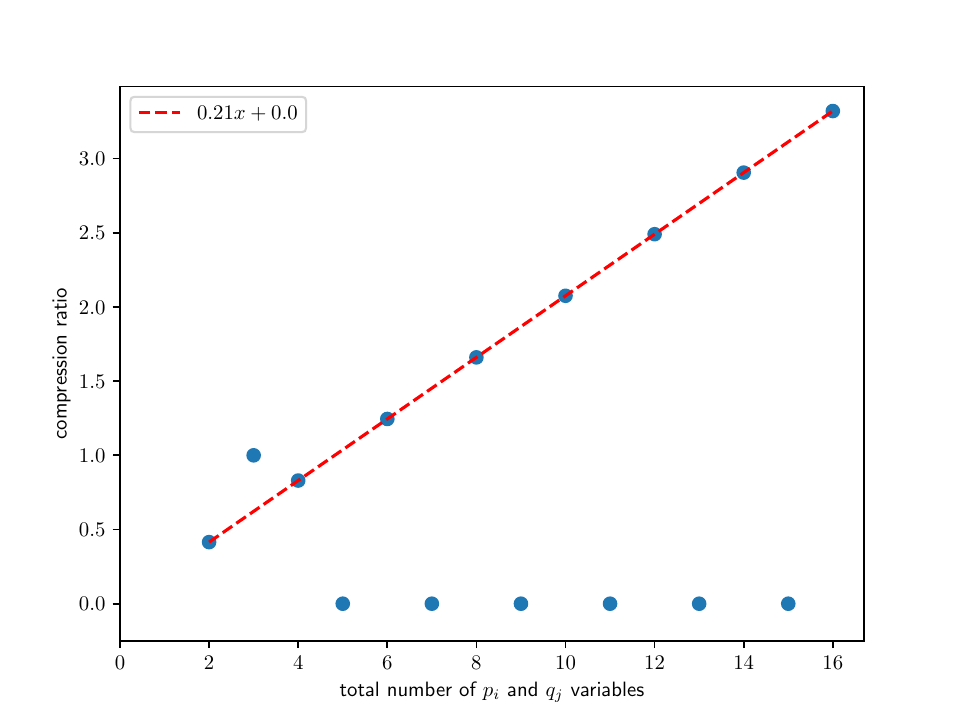}
        \caption{$m = 714433477$}
        \label{subfig:image2}
    \end{subfigure}
    
    \begin{subfigure}{0.45\textwidth}
        \includegraphics[width=\linewidth]{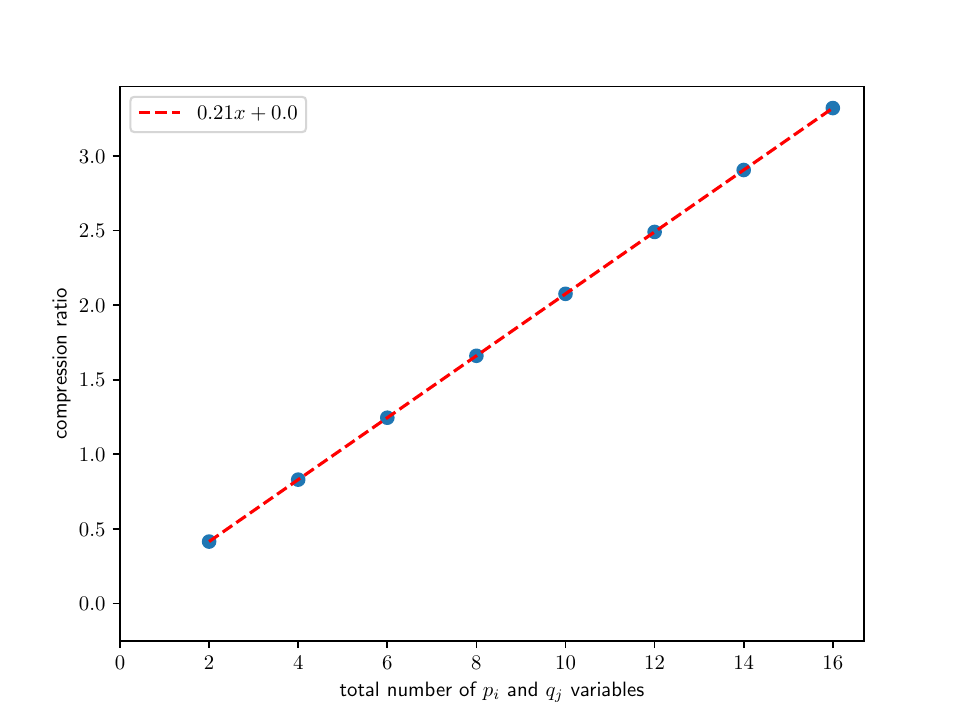}
        \caption{$m = 1753778247857$}
        \label{subfig:image3}
    \end{subfigure}
    \hfill
    \begin{subfigure}{0.45\textwidth}
        \includegraphics[width=\linewidth]{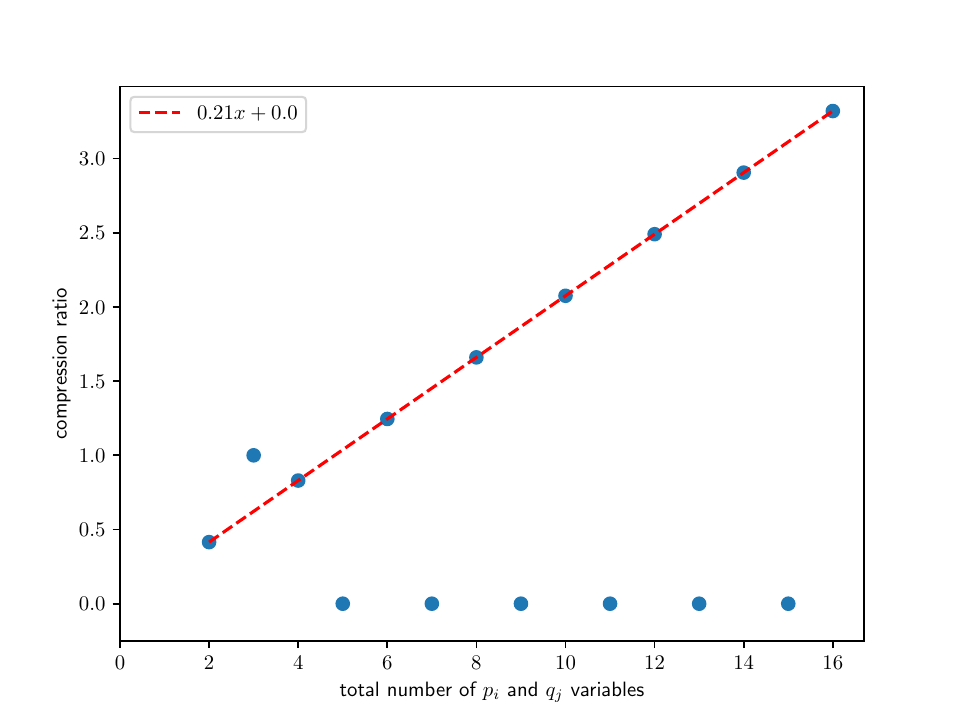}
        \caption{$m = 903873756190981$}
        \label{subfig:image4}
    \end{subfigure}
    
    \caption{Compression ratios for different total numbers of $p_i$ and $q_j$ variables in clauses.}
    \label{fig:compression_ratios}
\end{figure}

In addition, we considered how much the search space can be reduced for clauses with a particular number of $p_i$ and $q_j$ variables. The results are presented in  Fig.~\ref{fig:compression_ratios}. We generated clauses using the classical preprocessing rules as proposed in~\cite{anschuetz2019variational} and we identified those for which the total number of $p_i$ and $q_j$ variables appearing in a clause was at most 16. We can see that the compression ratio, defined as the logarithm of the number of all assignments minus number of invalid assignments over $p_i,q_j$ grows at most linearly with the total number of $p_i$ and $q_j$ variables, however it does not change with length of the biprime. This is because the clauses with chosen number of variables are mostly those specifying conditions for the least and the most significant bits of the product, and they often have a similar form.
% Since we do not provide any good candidate for the certification for the Func-QAOA, we also left how to construct a good initial states as an open question.

Note that the origin of finding such clauses does not have to be directly connected to the certificate used. In particular all the rules generated for the smaller prime based on the integer formulation for multiplication can be also used for the modulo-based certification. Note that the simplification can be done only based on some of the constraints, and the full formulation does not need to be constructed.  

It is important to acknowledge that this simplification is unlikely to reduce the search space to $ \order{\log n}$ qubits. These kinds of simplifications are classical in nature and one would expect the same simplification to occur if the same problem is tackled using a classical algorithm e.g. simulated annealing, which would make the problem classically tractable. 

\section{Discussion and Conclusion} \label{sec:conclusions}

In this paper, we pointed some potential challenges that may arise when attempting to solve the factorization problem with QAOA. This includes in particular representing factorization problem as a SAT instance, involving bits that are redundant or even lack context which occur as a result of formulating the problem instance as a QUBO or HOBO, and using a bit-oriented mixer. We showed, how those limitations can be potentially overcame by leveraging the results from the current literature, with a particular focus on Func-QAOA, a variant of QAOA which relies on the universality of the gate-based model. With Func-QAOA, not only the context of the problem remains visible in the QAOA ansatz, but it also allows a natural incorporation of more complicated certifications for the problem.

While our work introduces a potentially interesting framework for integer factorization using QAOA, it remains unclear whether a practical certificate exists within the limitations of the NISQ era. Among all the certificates we provided and are aware of, the ones that offer subexponential speed-up for classical computing require subexponential time and qubits for certification. This clearly goes beyond the capabilities of current devices, as achieving sufficiently faithful evolution with subexponentially many gates would likely require fidelity that would allow for running Shor's algorithm. Therefore, we leave an open question whether there is a certificate which has subexponential probability of finding a correct solution, with a polynomial number of gates. While one might hope that exploring the search space could eliminate the requirement for subexponential success probability, this would imply that QAOA achieves faster-than-polynomial speedup for that optimization problem, which we believe to be highly unlikely.

We would also like to emphasize that the challenge of finding a good certificate is not the only aspect that needs to be addressed. Despite the promising results on applying QAOA for a \emph{decision problem} $K$-SAT~\cite{boulebnane2022solving}, it is far from evident whether similar results can be expected for factorization. Furthermore, there are crucial factors that have not been taken into account, but will contribute to a big constant, and likely at least polynomial multiplicative overhead for the time of finding the correct solution. These factors include the limited connectivity of the quantum devices, the impact of noise and decoherence, the cost of estimating the energy of the quantum state, and the cost of the optimization process itself. Taking all of those into account, and the unpromising recent results for QAOA on real quantum machines~\cite{pelofske2023quantum}, we assert that, at the very least, in the near future, QAOA is unlikely to pose a threat to the security of our bank accounts.

\section*{Acknowledgements}

\"O.S. and L.B. acknowledge support from the National Science Center under grant agreement 2019/33/B/ST6/02011. A.G. has been partially supported by the National Science Center under grant agreements 2019/33/B/ST6/02011 and 2020/37/N/ST6/02220. We would like to thank Andris Ambainis for the discussion on the quantum factorization methods. We acknowledge QWorld Association for organizing the remote internship program QIntern 2022, during which we initiated the work presented in this paper. 

% \section*{Author contributions statement}

% A.G. and \"O.S. proposed the investigation topic and quantified its impact. \todo{do not forget about the rest} All authors contributing to preparation of the paper.

\section*{Additional information}

Code used for generating Figs.~\ref{fig:space-reduction-scheme} and~\ref{fig:compression_ratios} is available at \url{https://doi.org/10.5281/zenodo.8363223}.

% Authors declare no competing interests.

\bibliographystyle{ieeetr}
\bibliography{factorization}

\end{document}